\documentclass[aps,preprint,superscriptaddress]{revtex4-1}
\usepackage{graphicx}
\usepackage{dcolumn}
\usepackage{amssymb}
\usepackage{amsmath}
\usepackage{bm}
\usepackage{pifont}
\usepackage{epsfig}
\usepackage{psfrag}
\usepackage[usenames]{color}
\begin{document}
\title{Direct Evidence for an Absorbing Phase Transition Governing Yielding of a Soft Glass}

\author{K. Hima Nagamanasa $^{\ast}$}
\affiliation{ Chemistry and Physics of Materials Unit, Jawaharlal Nehru Centre for Advanced Scientific Research, Jakkur, Bangalore - 560064, INDIA}
\author{Shreyas Gokhale}
\affiliation{Department of Physics, Indian Institute of Science, Bangalore - 560012, INDIA}
\author{A. K. Sood}
\affiliation{Department of Physics, Indian Institute of Science, Bangalore - 560012, INDIA}
\affiliation{International Centre for Materials Science, Jawaharlal Nehru Centre for Advanced Scientific Research, Jakkur, Bangalore - 560064, INDIA}
\author{Rajesh Ganapathy $^{\ast}$}
\affiliation{International Centre for Materials Science, Jawaharlal Nehru Centre for Advanced Scientific Research, Jakkur, Bangalore - 560064, INDIA}
\date{\today}
\draft

\begin{abstract}
We present the first direct experimental evidence showing that yielding of a prototypical soft solid - a colloidal glass -  is a non-equilibrium `absorbing' phase transition. By simultaneously quantifying single-particle dynamics and bulk mechanical response, we extracted critical exponents for the order parameter and the relaxation time and found that this transition belongs to the conserved directed percolation universality class. In addition, the observed critical slowing down is accompanied by a growing correlation length associated with the size of regions of high Debye-Waller factor which are precursors to yield events in glasses. Our results show unambiguously that yielding of soft solids falls squarely in the realm of non-equilibrium critical phenomena.
\end{abstract}

\maketitle
\renewcommand{\thefootnote}

A wide variety of solids including atomic crystals, metallic glasses, dense suspensions, gels and foams exhibit yielding and plastic flow when subjected to sufficiently large external stresses \cite{miguel2006jamming,barnes1999yield}. Apart from representing an interesting class of non-equilibrium processes, these phenomena are routinely exploited in numerous industrial applications \cite{valiev2004nanostructuring}. Elucidating mechanisms that govern yielding and plasticity therefore assumes fundamental as well as technological significance. In hard crystalline materials, yielding proceeds via the motion of well-defined topological defects called dislocations. While no such topological defects exist in amorphous solids, flipping of shear transformation zones is thought to be responsible for plastic deformation. In both crystalline as well as amorphous solids, however, it is now well-established that microscopic irreversible yield events during plastic deformation occur collectively as intermittent avalanches with power-law size distribution \cite{dimiduk2006scale,miguel2001intermittent,sun2010plasticity}. Further, the observation of self-organized criticality in these systems suggests an underlying non-equilibrium phase transition \cite{PhysRevLett.59.381,PhysRevA.38.364}, which theories and simulations posit to be centered at the yield point \cite{tsekenis2013determination}. Together, these observations indicate the emergence of robust dynamical features near yielding that are insensitive to microscopic details. In the context of soft solids by contrast, such an holistic understanding of yielding is yet to emerge. On the experimental front, light scattering echo measurements have identified a strain threshold for the onset of irreversible rearrangements \cite{petekidis2002rearrangements}, but it is unclear whether it corresponds to the yield point. Moreover, light scattering techniques are not particularly well-suited to probe the collective nature of local plastic events. The existence of such cooperativity has been inferred indirectly from observations of long-ranged strain correlations in colloidal glasses \cite{chikkadi2011long} and the power-law scaling of dislocation velocities in colloidal crystals \cite{pertsinidis2005video}. However, despite recent efforts \cite{keim2013yielding,mohan2013local}, whether these correlations lead to a phase transition still remains an open question.

In this Letter, by combining particle scale imaging with bulk rheology we show that the transition from reversible to irreversible dynamics during yielding of a prototypical soft solid - a binary colloidal glass - is associated with a non-equilibrium critical point. Thus far, experiments have been unable to uncover the existence of such a critical phenomenon mainly because they employed steady shear, where the continuous accumulation of strain precludes the characterization of critical behavior near yielding \cite{divoux2010transient}. Here, the application of oscillatory shear allowed us to not only identify the onset of irreversibility but also extract critical exponents. Intriguingly, the same exponents have also been observed in periodically driven dilute non-Brownian suspensions \cite{corte2008random}. However, the critical phenomenon seen here has a fundamentally different origin. The most important distinction is that our particles are Brownian and owing to thermal fluctuations, the critical strain for the onset of irreversibility is identically zero in the dilute limit. It is only in sufficiently dense Brownian suspensions, where particle caging by neighbors results in finite rigidity, that the transition to irreversibility occurs at a non-zero strain \cite{petekidis2002rearrangements}. Together, these features of Brownian suspensions enabled us to directly link the observed phase transition with the yield point. In a broader context, since colloids are convenient model systems to probe atomistic phenomena \cite{schall2007structural,wang2012imaging}, insights gleaned from the present studies should be relevant to hard materials as well.

Our glass comprised of an aqueous suspension of fluorescently labelled temperature-sensitive size-tunable poly N-isopropylacryl amide (PNIPAM) colloidal spheres of radii of 1 and 2 $\mu$m (see Supplemental Material \cite{supplementary}). Apart from exhibiting structural disorder like most soft solids, our system is also amenable to 3D single-particle resolution imaging. The volume fraction used $\phi \approx 0.67$ was beyond the glass transition $\phi_{g} \sim 0.58$. A number density ratio of 1:3 of large and small particles adequately suppressed crystallization, as confirmed from 3D confocal imaging. A commercial rheometer mounted on a fast confocal microscope allowed us to investigate the particle scale dynamics of the glass under shear (see Supplemental Material \cite{supplementary}). The bulk elastic and viscous moduli $G'$ and $G''$, respectively, were measured by applying oscillatory strains $\gamma = \gamma_{o}\sin\omega t$, where $\gamma_{o}$ is the strain amplitude and $\omega$ is the frequency. Subsequently, we measured $G'$ and $G''$ as a function of the oscillation cycle number $\tau$ for various $\gamma_{o}$'s at $\omega = 1$ rad/s. For each $\gamma_{o}$, single-particle dynamics was quantified by imaging a 54 $\mu$m$\times$54 $\mu$m 2D slice oriented parallel to the velocity-vorticity plane (see Supplemental Material \cite{supplementary} Fig. S1) and located 7 $\mu$m away from the fixed bottom plate, at 60 frames per second. To visualize the 3D microstructure, we obtained image stacks (25 $\mu$m$\times$25 $\mu$m$\times$8 $\mu$m) along the velocity gradient direction at a few select $\gamma_{o}$'s.

\begin{figure}[tbp]
\includegraphics[width=0.7\textwidth]{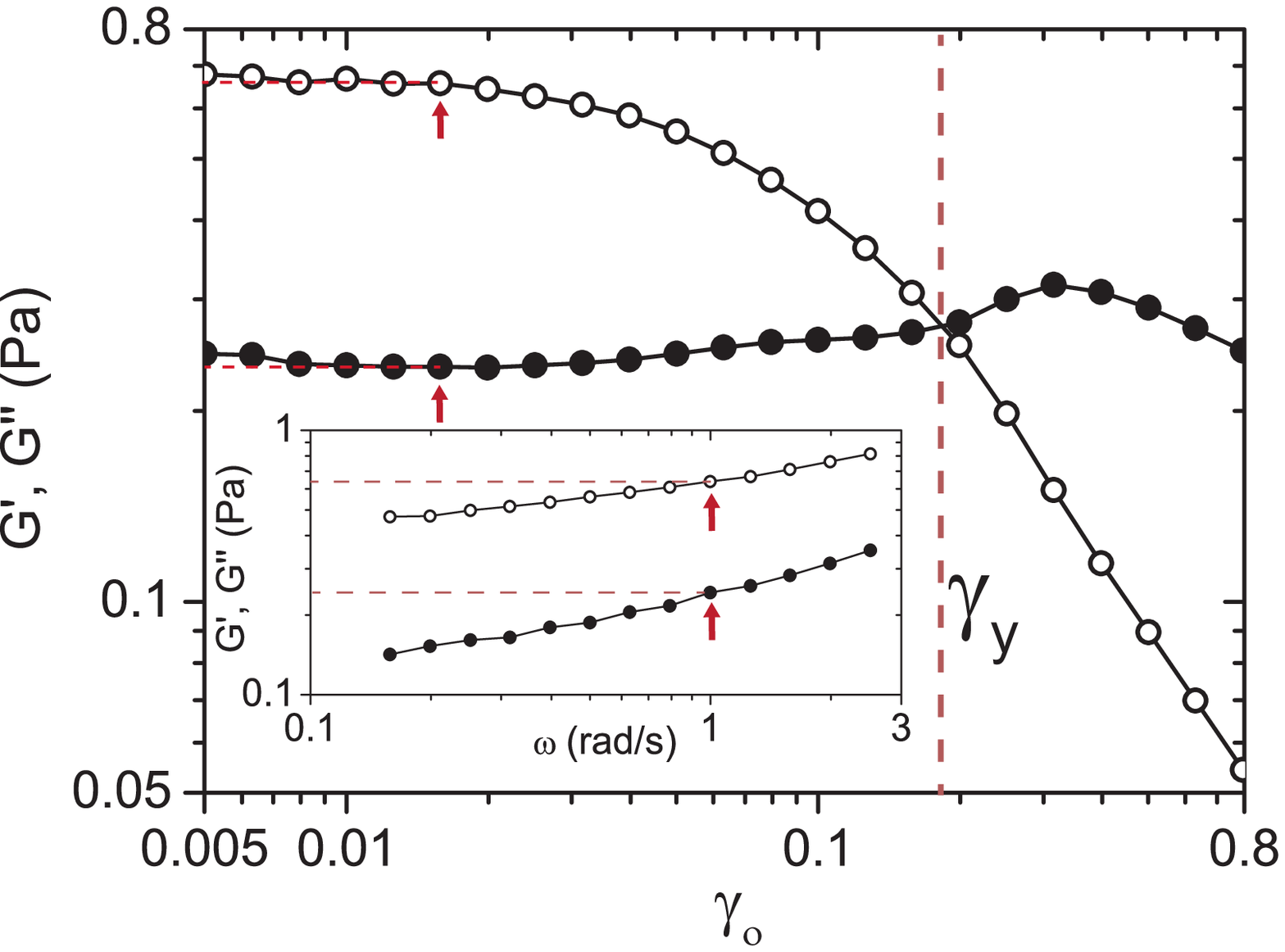}
\caption{$\gamma_{o}$-sweep measurements performed at $\omega$ = 1 rad/s. Inset shows results from $\omega$-sweep experiments at $\gamma_{o} = 0.015$. $G'$ and $G''$ are denoted by ({$\boldsymbol \circ$}) and ({$\boldsymbol \bullet$}), respectively. The red arrows and dotted lines in the figure and the inset highlight the values of $G'$ and $G''$ for $\gamma_{o}$ = 0.015 and $\omega$ = 1 rad/s. }
\label{Figure1}
\end{figure}

The rheological response of our sample is typical of a soft glass (Fig. \ref{Figure1}) \cite{sollich1998rheological,carrier2009nonlinear,*erwin2010dynamics,*rogers2011oscillatory,*rogers2011sequence,van2013rheology}. We found only a weak dependence of $G'$ and $G''$ on $\omega$, which independently confirmed that our samples were in the glassy state (inset to Fig. \ref{Figure1}) \cite{sollich1998rheological}. From $\gamma_{o}$-sweep experiments (Fig. \ref{Figure1}), the $\gamma_{o}$ corresponding to the crossover of $G'$ and $G''$ was identified as the yield strain $\gamma_{y} = 0.2$ \cite{van2013rheology}. The ubiquitous $G''$ peak was located beyond $\gamma_{y}$ (Fig. \ref{Figure1}), which along with the high P\'eclet number of our system (see Supplemental Material \cite{supplementary}), suggests that yielding in our system may be governed by shear-induced rearrangements rather than thermally assisted cage jumps \cite{koumakis2013complex}. 

Next, we characterized the irreversible micro-structural changes that lead to yielding. A natural measure of irreversibility is the fraction, $f_{IR}$, of particles that do not return to their initial positions at the end of a strain cycle \cite{corte2008random} (see Supplemental Material \cite{supplementary}). Since $f_{IR}$ is independent of the exact nature of local yield events, it should be generically applicable to ordered as well as disordered soft solids. Following pre-shear, the relaxation of residual stresses during subsequent strain cycles should result in irreversible changes in particle configurations \cite{keim2013yielding}. We therefore expect $f_{IR}$ to evolve in time towards a steady state value. Confocal volumes highlighting irreversible events show that $f_{IR}$ exhibits a pronounced increase across $\gamma_{y}$ \cite{lundberg2008reversible} (see Supplemental Material \cite{supplementary} Fig. S3 ). Further, $f_{IR}$ predominantly comprises of small particles due to their lower strain threshold for irreversible rearrangements. 

\begin{figure}[htbp]
\includegraphics[width=1\textwidth]{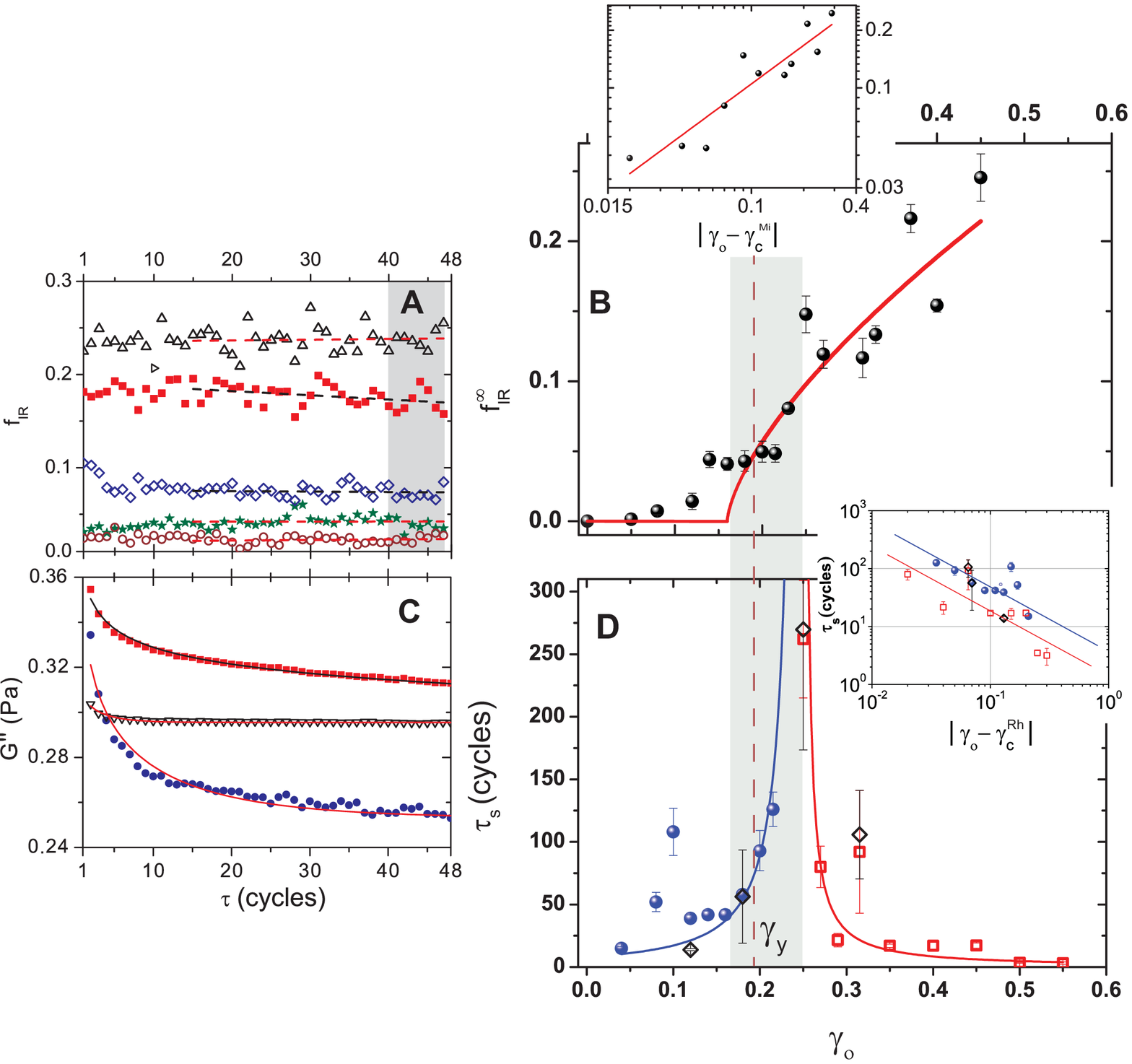}
\caption{(A) $f_{IR}(\tau)$ for $\gamma_{o} = 0.05$ ({\color{Brown} $\boldsymbol \circ$}), $\gamma_{o} = 0.12$ ({\color{Green} $\boldsymbol \bigstar$}), $\gamma_{o} = 0.215$ ({\color{Blue} $\boldsymbol \diamond$}), $\gamma_{o} = 0.25$ ({\color{Red} $\boldsymbol \blacksquare$}) and $\gamma_{o} = 0.37$ ({\color{Black} $\boldsymbol \triangle$}). The dashed lines are linear fits to the data.  (B) $f_{IR}^{\infty}$ as a function of $\gamma_{o}$. Inset to B shows $f_{IR}^{\infty}$ versus $|\gamma_{o} - \gamma_{c}^{Mi}|$. $f_{IR}^{\infty}(\gamma_{o}) =  f_{IR}^{ss}(\gamma_{o}) -  f_{IR}(\gamma_{o} = 0) $, where $f_{IR}^{ss}$ is the steady state fraction of irreversible rearrangements obtained by averaging over the shaded region in A. $f_{IR}(\gamma_{o} = 0)$ = 0.023. The red curve is a power-law fit to the data. $\gamma_{c}^{Mi}$ and $\beta$ were extracted by minimizing $\chi^2$. (C) $G''(\tau)$ for $\gamma_{o} = 0.04$ ({\color{Blue} $\boldsymbol \bullet$}), $\gamma_{o} = 0.25$ ({\color{Red} $\boldsymbol \blacksquare$}) and $\gamma_{o} = 0.45$ ({\color{Black} $\boldsymbol \triangledown$}). The curves are best fits to the data. (D) Relaxation time $\tau _{s}$ versus $\gamma_{o}$.  The solid curves represent power-law fits to the data. Inset to D shows $\tau _{s}$ versus $|\gamma_{o} - \gamma_{c}^{Rh}|$ . The red line is a fit to the data whereas the blue line is a guide to the eye and has the same slope as the red line. ({\color{Blue} $\boldsymbol \bullet$}) and ({\color{Red} $\boldsymbol \square$}) correspond to the regimes $\gamma_{o} < 0.25$ and $\gamma_{o} > 0.25$, respectively. Black Diamonds ({\color{Black} $\boldsymbol \diamond$}) correspond to $\tau _{s}$ obtained from independent measurements on the same sample.}
\label{Figure2}
\end{figure}

Owing to the poor temporal resolution in 3D imaging at large $\gamma_{o}$'s, we resorted to 2D to quantify $f_{IR}(\tau)$ for all $\gamma_{o}$'s investigated. Here, particles were labelled irreversible if, apart from meeting the displacement threshold, they lost at least four nearest-neighbors at the end of a cycle (see Supplemental Material \cite{supplementary}). Studies on supercooled liquids and glasses have shown that configurational changes ensuing from such irreversible dynamics are permanent \cite{widmer2008irreversible}. For $\gamma_{o}$'s far from $\gamma_{y}$, we did not observe any transients in $f_{IR}(\tau)$ implying that steady state was reached rapidly (Fig. \ref{Figure2}A). Near $\gamma_{y}$, for $\gamma_{o}=0.25$ however, $f_{IR}(\tau)$ shows a slow fall over the experimental duration. Similar long-lived transients near yielding have also been observed in recent simulations of periodically sheared amorphous solids \cite{fiocco2013oscillatory,*regev2013onset}. To investigate the onset and $\gamma_{o}$-dependence of irreversibility, we plotted the steady state fraction of irreversible particles after subtracting the thermal contribution, denoted by $f_{IR}^{\infty}$, as a function of $\gamma_{o}$ in  Fig. \ref{Figure2}B (also see Supplemental Material \cite{supplementary} for Movie S1). $f_{IR}^{\infty}$ remains small up to a critical strain $\gamma_{c}^{Mi}$ and increases rapidly beyond $\gamma_{c}^{Mi}$ \cite{petekidis2002rearrangements}. Here, the superscript `${Mi}$' signifies that the critical strain has been extracted from microscopy data. Further, $f_{IR}^{\infty}$ exhibits power-law scaling of the form $f_{IR}^{\infty} \propto (\gamma_{o} - \gamma_{c}^{Mi})^{\beta}$ for $\gamma_{o} > \gamma_{c}^{Mi}$, suggesting order parameter-like behaviour. The best-fit power-law gave $\beta = 0.67 \pm 0.09$ and a $\gamma_{c}^{Mi}$ of $0.16$, which is remarkably close to $\gamma_{y}$ (Inset to \ref{Figure2}B).

The dependence of $f_{IR}^{\infty}$ on $\gamma_{o}$ strongly suggests a non-equilibrium phase transition governing yielding and the long-lived transients in $f_{IR}(\tau)$ likely correspond to critical slowing down. Since irreversible dynamics leads to energy dissipation, the transients in $f_{IR}$ are also evident in $G''(\tau)$  (Fig. \ref{Figure2}C). Unlike $f_{IR}(\tau)$, $G''(\tau)$ is a bulk measure and therefore has a significantly better signal-to-noise ratio. Analogous to \cite{corte2008random}, the relaxation curves in Fig. \ref{Figure2}C are well-fitted by the functional form $G''(\tau) = (G''_{o} - G''_{\infty}){e^{-\tau \slash \tau_{s}} \over \tau^{\delta}} + G''_{\infty}$, where $\tau_{s}$ is the time taken to reach steady state and $G''_{o}$ and $G''_{\infty}$ are the initial and steady state values of $G''(\tau)$, respectively. This functional form captures the crossover from exponential to power-law behaviour expected near a critical point. Remarkably, $\tau_{s}$ extracted from the fits diverges as $\tau_{s} \propto |\gamma_{o} - \gamma_{c}^{Rh}|^{-\alpha}$, with $\alpha = 1.1 \pm 0.3$, where $\gamma_{c}^{Rh}$ denotes the critical strain extracted from rheology data. We find that $\gamma_{c}^{Rh} = 0.25$ a value close to $\gamma_{y}$ (Fig. \ref{Figure2}D and Inset. Also see Supplemental Material \cite{supplementary}). Collectively, these observations (Fig. \ref{Figure2}B \& D) provide strong evidence for a non-equilibrium phase transition at the yield point. 

Yielding in hard materials has often been associated with the pinning-depinning transition \cite{tsekenis2013determination}. However, quenched disorder, which is an important ingredient of pinning-depinning transitions, is absent in our system. The $\gamma_{o}$ dependence of $f_{IR}^{\infty}$ observed here suggests a transition from reversible dynamics with $f_{IR}^{\infty} \approx 0$, to irreversible dynamics with $f_{IR}^{\infty} > 0$, which is characteristic of an absorbing phase transition (APT) \cite{corte2008random}. Most APTs belong to the directed percolation universality class \cite{hinrichsen2000non}. However, the particle number density being conserved in our system, we expect the non-equilibrium phase transition seen here to belong to the conserved directed percolation (C-DP) universality class \cite{menon2009universality}. Further, in our cone-plate shear geometry, the diameter of the cone is $\sim 200$ times larger than the gap and hence, the exponents extracted here should be in agreement with those observed for 2D C-DP \cite{menon2009universality}. Indeed, we find that the relaxation time exponent $\alpha = 1.1 \pm 0.3$ (Fig. \ref{Figure2}D) and the order parameter exponent $\beta = 0.67 \pm 0.09$ (Fig. \ref{Figure2}B) are very close to the C-DP universality class value in 2D \cite{menon2009universality}. This consistency in exponents extracted from simultaneous bulk rheological and single-particle measurements reflects the robustness of the underlying phase transition.

\begin{figure}[tbp]
\includegraphics[width=0.5\textwidth]{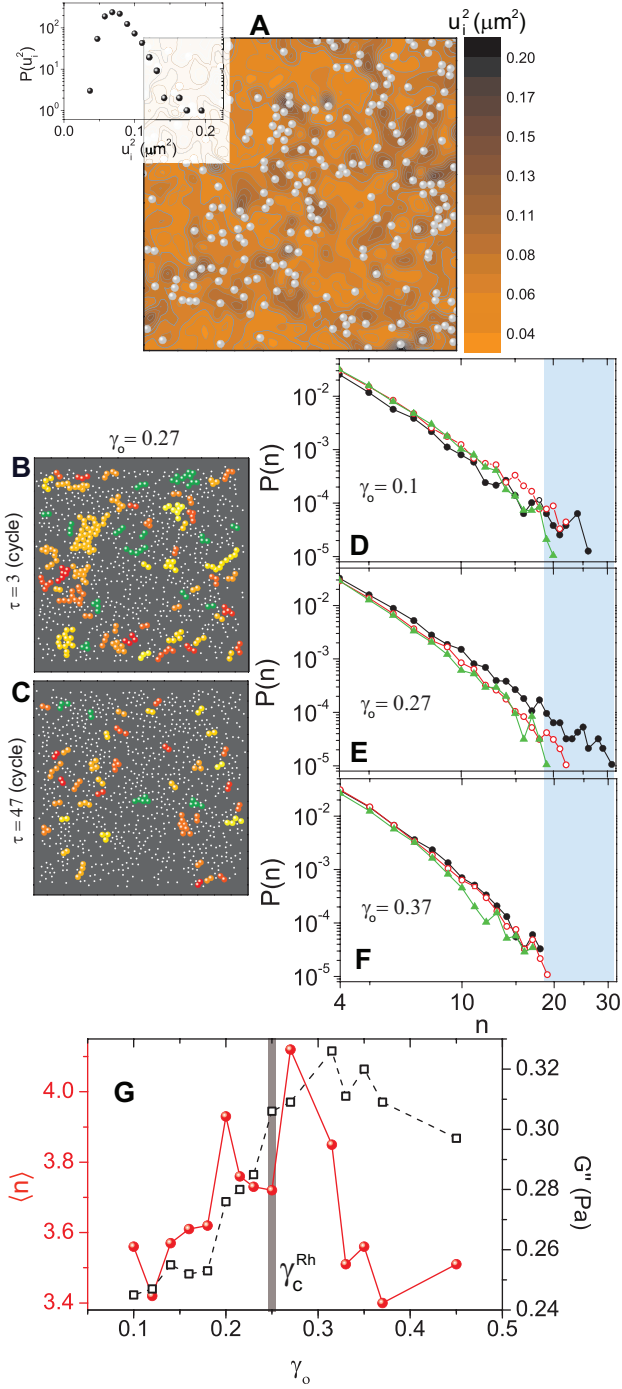}
\caption{(A) Color map of $u_{i}$ for $\tau = 2$. The solid spheres represent irreversible particles at the end of the same cycle. Inset shows the distribution of $u_{i}$. (B)-(C) Representative snapshots with the top 10$\%$ high $u_{i}$ particles shown as big solid spheres and the remaining shown as small circles. The colors are a visual aid to help demarcate clusters. (D)-(F) Distribution of cluster size $P(n)$. In (D)-(F) P(n) for $\tau = 2$ to $10$ ({$\boldsymbol \bullet$}), for $\tau = 20$ to $30$ ({\color{Red} $\boldsymbol \circ$}) and for $\tau = 37$ to $47$ ({\color{Green} $\boldsymbol \blacktriangle$}). (G) $\langle n \rangle$ as a function of $\gamma_{o}$ is shown as ({\color{Red} $\boldsymbol \bullet$}). $G''$ from bulk rheology is shown as ({$\boldsymbol \square$}).}  
\label{Figure3}
\end{figure}

The observation of critical slowing down implies increasing spatial correlations between local irreversible rearrangements near yielding. In our system, these correlations can be probed using concepts developed to understand the dynamics of amorphous solids. Recent simulations suggest that local plastic events in glasses originate from spatially localized low frequency vibrational modes \cite{tanguy2010vibrational}. Since these localised regions also posses low stiffness values, we identified them using a local measure of elasticity, namely, the Debye-Waller factor $u_{i}$ \cite{tsamados2009local} defined as $u_{i} = (r_{i} - \langle r_{i}\rangle)^{2}$ \cite{widmer2006predicting}. Here, $r_{i}$ is the instantaneous position of particle $i$, and $\langle r_{i}\rangle$ is its mean position averaged over a suitable time window (see Supplemental Material \cite{supplementary}). Experimentally, $u_{i}$ is a fairly accurate measure since it requires particles to be tracked only for a short duration. Figure \ref{Figure3}A shows a color map of $u_{i}$ averaged over an oscillation cycle, for $\gamma_{o} = 0.25$ and $\tau = 2$ and the inset shows its distribution $P(u_{i})$. We find that regions of high $u_{i}$ are spatially localised and correlated with subsequent irreversible rearrangements (orange circles in Fig. \ref{Figure3}A). The transient dynamics of $f_{IR}(\tau)$ (Fig. \ref{Figure2}A), should therefore stem from the spatio-temporal evolution of these high $u_{i}$ regions. To show this, we first identified the top 10\% high $u_{i}$ particles for various $\gamma_{o}$'s. Figure \ref{Figure3}B \& C show snapshots of these particles for $\gamma_{o} = 0.27$ at $\tau = 2$ and $47$, respectively. The top 10\% high $u_{i}$ particles are spatially clustered and more importantly, there are fewer large clusters for $\tau = 47$. Figure \ref{Figure3}D-F shows the cluster size distribution $P(n)$, where $n$ is the number of particles in a cluster, for three different time intervals at various $\gamma_{o}$'s. Analogous to $f_{IR}(\tau)$, $P(n)$ is stationary for $\gamma_{o}$ far from $\gamma_{c}^{Rh}$ and evolves steadily in the vicinity of $\gamma_{c}^{Rh}$. Further, near $\gamma_{c}^{Rh}$, $P(n)$ for $n > 20$ is significantly smaller at a larger $\tau$. It is also evident that on average, clusters are larger for $\gamma_{o} \approx \gamma_{c}^{Rh}$ (Fig. \ref{Figure3}E) as compared to $\gamma_{o}$ far away from $\gamma_{c}^{Rh}$ (Fig. \ref{Figure3}D \& F). Indeed, the average cluster size $\langle n \rangle = {\sum n^{2}P(n) \over \sum nP(n)}$, a commonly used measure of the correlation length \cite{weeks2000three}, shows a clear maximum near $\gamma_{c}^{Rh}$ (Fig. \ref{Figure3}G. Also see Supplemental Material \cite{supplementary}) which signals the increasing spatial correlation between local irreversible rearrangements.

Our results demonstrate that correlations between local yield events not captured by mean field theories like Soft Glassy Rheology (SGR) \cite{sollich1997rheology} lead to an APT at the yield point. For $\gamma_{o}$ less than a critical strain, yield events are rare and the system quickly self-organizes into an `absorbing' steady state ($f_{IR}^{\infty} \approx 0$), where the applied strain is insufficient to induce irreversible changes in particle configuration. For $\gamma_{o}$ larger than the critical strain, the imposed strain facilitates many independent irreversible rearrangements, correlations between yield events are washed out, and the system rapidly reaches a `fluctuating' steady state ($f_{IR}^{\infty} > 0$) \cite{sollich1997rheology}. Close to the critical strain, however, correlations between local yield events trigger a cascade of irreversible rearrangements, which is manifested as a growing length scale (Fig. \ref{Figure3}G) and leads to critical slowing down (Fig. \ref{Figure2}D). Further, it has not skipped our attention that the growing cluster sizes near $\gamma_{y}$ may also have implications for the origin of the $G''$ peak (Fig. \ref{Figure3}G). This is especially important when the $G''$ peak coincides with the yield strain, a scenario frequently observed in soft solids. 

To summarize, through simultaneous quantification of single-particle dynamics and bulk viscoelastic moduli, we have discovered a non-equilibrium critical phenomenon governing yielding of a colloidal glass (Fig. \ref{Figure2}B \& D). Unlike non-Brownian suspensions, our Brownian system has a finite threshold for irreversibility only at sufficiently large volume fractions where it behaves as a viscoelastic solid. This allowed us to unambiguously identify the observed critical point with the yield strain. We found that the growing time scale in our experiments is accompanied by a growing length scale associated with clusters of particles with high Debye-Waller factor (Fig. \ref{Figure3}G), which are precursors of local plastic events in amorphous solids (Fig. \ref{Figure3}A). We therefore expect the correlations between local irreversible rearrangements observed here to have a correspondence with observations of avalanches in sheared amorphous solids \cite{lemaitre2009rate}. Given that a wide range of soft materials such as gels, emulsions and foams also exhibit strikingly similar rheological properties, the phase transition observed here should be generic to soft solids. Further, since the mechanical response of soft solids depends on the frequency, albeit weakly (see Supplemental Material \cite{supplementary} Fig. S2 ), it would also be fascinating to examine the frequency dependence of the onset of irreversibility in these materials. Our results exemplify the need for refining mean field theories like SGR. In particular, it might be possible to incorporate explicit interactions in SGR such that the local yielding of one element assists in the yielding of a neighbouring element. Connections of such an extended SGR model to theories that relate yielding to the percolation of a liquid phase within a deformed solid \cite{liu2013quasi} may also be conceivable. Most importantly, our findings set the stage for developing a unified framework for yielding of soft solids.

The authors thank Michael Cates, Jack Douglas, Sriram Ramaswamy, Narayanan Menon and Rema Krishnaswamy for helpful discussions. K.H.N. thanks the Council for Scientific and Industrial Research (CSIR) India for a Senior Research Fellowship. S.G. thanks CSIR India for a Shyama Prasad Mukherjee Fellowship. A.K.S. thanks Department of Science and Technology, India for support under J.C. Bose Fellowship and R.G. thanks the International Centre for Materials Science (ICMS) and the Jawaharlal Nehru Centre for Advanced Scientific Research (JNCASR) for financial support.

\subsection*{Author Contributions}
K.H.N. and S.G. contributed equally to this work.\\
$^{\ast}$ Corresponding author

\bibliography{references} 
\bibliographystyle{apsrev4-1} 

\newpage

\end{document}